# The Proceeding of First Work-in-Progress Session of
## The CSI International Symposium on Real-Time and Embedded Systems and Technologies (WiP-RTEST 2018)
### University Of Tehran
### May 9-10

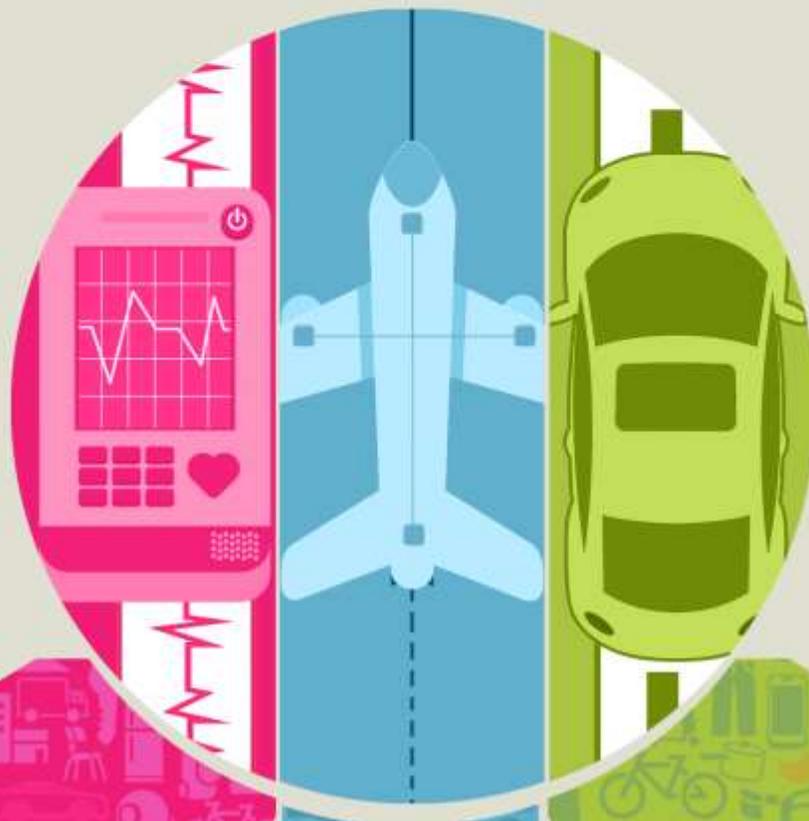

2018.RTEST-conf.org

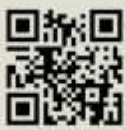

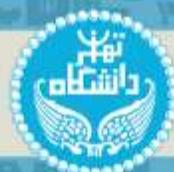

# The Proceeding of First Work-in-Progress Session of 2018 CSI International Symposium on Real-Time and Embedded Systems and Technologies (WiP-RTEST 2018)

May 9-10, 2018

University of Tehran, Iran

# Table of Contents

# Preface

Welcome to the 2[nd] CSI International Symposium on Real-Time and Embedded Systems and Technologies (RTEST), and first Work-in-Progress (WiP) session in the University of Tehran (College of Engineering), Tehran, Iran, May 9-10, 2018.

The present volume contains the proceedings of RTEST WiP 2018, chaired by Marco Caccamo, University of Illinois at Urbana-Champaign. This event has been organized by the School of Electrical and Computer Engineering at the University of Tehran, in conjunction with the Department of Computer Engineering at Sharif University of Technology, Tehran, Iran. The topics of interest in RTEST WiP span over all theoretical and application-oriented aspects, reporting design, analysis, implementation, evaluation, and empirical results, of real-time and embedded systems, internet-of-things, and cyber-physical systems. The program committee of RTEST 2018 consists of 54 top researchers in the mentioned fields from top universities, industries, and research centers around the world.

RTEST 2018 has received a total of 41 submissions, out of which we have accepted 14 regular papers and 4 work-in-progress papers. Each submission has been reviewed by 3 to 5 independent referees, for its quality, originality, contribution, clarity of presentation, and relevance to the symposium topics.

Two distinguished keynote speakers, Radu Grosu and Sahra Sedigh Sarvestani, have been invited to deliver their lectures at RTEST 2018.

We thank the computer society of Iran, IEEE and IEEE Iran section, University of Tehran, Sharif University of Technology, and all the financial sponsors listed in the pre-proceedings for their support of all kind. We also thank the members of organizing committee for organizing the event and all their time and efforts, the program committee for their time and excellent contributions, and the dean and associative heads of School of ECE at the University of Tehran to making RTEST a quality symposium.

We thank Fatemeh Ghassemi for her help in preparing this volume. Last but not least, our thanks go to the authors and symposium participants, without them RTEST would not have been possible.

May 9, 2018

Mehdi Kargahi

Alireza Ejlali

# Accelerating Multicore Scheduling in ChronOS Using Concurrent Data Structures

Ali Behnoudfar
School of Electrical and Computer Engineering
University of Tehran
Tehran, Iran
ali.behnoudfar@ut.ac.ir

Mehdi Kargahi
School of Electrical and Computer Engineering
University of Tehran
Tehran, Iran
kargahi@ut.ac.ir

*Abstract—* Real-time systems are increasingly coming to be implemented in multiprocessor and multicore platforms. In order to achieve full performance gain on these platforms, there is a need for efficient and scalable implementation. One possible source of inefficiency in these platforms is the shared data structure used for interaction and coordination between threads. In order to prevent race condition resulting from concurrent access to these shared data structures, a locking mechanism is usually used, which while providing safety, limits the performance gain, as at any time, data structure can be accessed by at most one thread of execution. Concurrent data structures try to address these issues. In this work shared data structure used in the context of a real-time multicore scheduling in a real-time operating system is changed to a concurrent version to achieve improved performance and scalability in these platforms.

## I. Introduction

In this section, the concepts used in this work are briefly explained.

### A. Concurrent Data Structures

Today multicore processors are used in anything from phones to laptops, desktops and servers. In these systems multiple threads of execution are executed concurrently which communicate with each other through shared memory. In order to achieve maximum speedup on these systems, it is necessary to have efficient and scalable concurrent implementation.

According to Amdahl's Law the amount of speedup gained by concurrent implementation of a program (executed on a multicore processor) is limited by amount of program executed sequentially. In many applications these sequential parts involve shared data structures, accessed concurrently by multiple threads of execution. Therefore, to increase speedup, it is worthwhile to derive as much parallelism as possible from these shared data structures using concurrent versions of them. [1]

The efficiency of these data structures is crucial to performance, but designing concurrent data structures is more difficult than sequential ones, because threads executing concurrently may interleave their steps in many ways, each with a different outcome.

### B. Real-Time Multicore Scheduling

Due to high performance requirements of real-time systems and prevalence of multiprocessor and multicore platforms in these systems, there is a need for efficient and optimized techniques for multicore scheduling on these systems [4].

Based on restrictions as to where tasks are allowed to execute, multicore scheduling algorithms fall into different categories: Global Scheduling allows jobs to migrate and execute on any processor, Partitioned Scheduling statically maps a task to a specific processor and afterwards jobs of those task can only execute on processor assigned to that task, Semi-Partitioned scheduling allows some degree of migration (for example a maximum of N migratory tasks in the system) and finally Clustered Scheduling partitions processors of a multiprocessor system into clusters and migration of tasks are allowed only within the cluster in which the task resides. [4].

One of the multicore real-time scheduling algorithms is global earliest deadline first (GEDF). In EDF scheduling algorithms, jobs are scheduled in order of increasing deadlines, with ties broken arbitrarily. Two multicore variants of this algorithm are Global Non-Preemptive EDF (G-NP-EDF) and Global Preemptive EDF (G-P-EDF) which are global scheduling algorithms. In G-NP-EDF, tasks can migrate, but once a job starts execution on a processor, it will run to completion on that processor without preemption. G-P-EDF allows jobs to be preempted and permits job migration with no restrictions. Figure 1 shows an example of these algorithms. Corresponding real-time properties is shown in Table 1 (task deadlines are equal to respective task periods). As we see in this figure none of these algorithms are optimal and deadlines can be missed in both (while total utilization is at most number of cores), task C misses it's deadline under G-P-EDF scheduling and task B misses it's deadline under G-NP-EDF scheduling.

| TASK | EXECUTION TIME | PERIOD |
|---|---|---|
| **A** | 1.5 | 3 |
| **B** | 2 | 3 |
| **C** | 4 | 6 |

Table 1. Real-time properties for figure 1

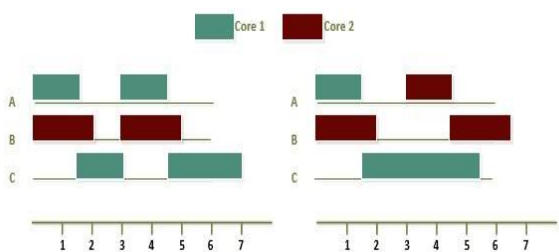

Figure 1. Left: G-P-EDF, right: G-NP-EDF

### C. ChronOS Real-Time Operating System

ChronOS Linux is a best-effort real-time Linux kernel for chip multiprocessors (CMPs). ChronOS was designed with following properties: a) real-time Linux kernel with the PREEMPT_RT patch (which enables preemption in kernel), b) providing
a scheduling framework for the implementation of a multicore real-time schedulers as plugins and c) OS support for multicore-aware real-time scheduling [2].
There are two ways for global scheduling in ChronOS:
1 - "Application Concurrent" scheduling: in this scheduling model, after the occurrence of a scheduling event (new task arrival, completion of a task, …) the global scheduler (such as G-NP-EDF in this model) which is running on the core on which mentioned event happened, picks a task for itself for execution.

2 - "Stop-the-World" (STW) scheduling: in this scheduling model, after the occurrence of a scheduling event, the global scheduler (such as G-P-EDF in this model) sends an "Inter-processor Interrupt" to all other cores after which execution on those cores is blocked until scheduler picks the tasks for all available cores [2]. "Application Concurrent" model is the focus of this work for reasons explained in section II.

### D. Flat Combining

Flat Combining (FC) is a synchronization paradigm based on coarse locking [3]. In summary in this technique, an execution thread tries to acquire a global lock on a sequential data structure, traverses all concurrent requests to it and then performs the combined requests on it.

## II. PROBLEM STATEMENT

In EDF scheduling, there is a need to find the task with the earliest deadline. For this purpose, usually a priority queue data structure is used (tasks are sorted by their deadline). Operations on these data structures consist of: insertion (arrival of a new task), removal (execution of a task is finished) and extractmin (retrieving the task with the smallest deadline). Consequently this data structure is accessed during all scheduling events and therefore is critical to the performance of the scheduling algorithm. In a highly concurrent execution, during all these scheduling events, there will be concurrent accesses to priority queue data structure, while it can be accessed by at most one thread at a time (due to being protected by a lock to prevent race condition). This situation could limit performance gain on a multicore processor considerably. Also due to high amount of extractmin requests, there will be a high amount of memory contention on the first element of priority queue. A concurrent priority queue improves performance in above scenario.
Since the explained issues only occur in "Application Concurrent" scheduling model in ChronOS (in "Stop-the-World" scheduling model, after a scheduling event execution on all other cores is blocked while scheduler picks tasks for them and consequently concurrent accesses that result in these issues will not be encountered), focus of this work will be on G-NP-EDF in "Application Concurrent" scheduling model in ChronOS.
To improve performance in above scenario, priority queue data structure used in G-NP-EDF scheduling algorithm in ChronOS, is changed to a concurrent version. The technique used to achieve this concurrency is FC. The two following properties of this technique made it appealing to be applied in this work:

1 – Synchronization overhead on frequently accessed shared locations (first element of priority queue) is reduced.

2 – Cache invalidation due to cache coherency is reduced as data structure is accessed by at most one thread at any time.

III. IMPLEMENTATION

As discussed in previous section, in this work priority queue data structure used in G-NP-EDF scheduling algorithm in ChronOS is changed to a concurrent version using FC technique.

ChronOS scheduler extends Linux scheduler. Each core has a queue of active tasks at each priority. Scheduler starts from highest priority and executes tasks in that order. For every priority there is another queue called ChronOS real-time run-queue (CRT-RQ) which stores pointers to ChronOS real-time tasks in the Linux run-queue. ChronOS real-time tasks start as Linux real-time tasks and then become ChronOS real-time tasks by entering a real-time segment. Additional data is added to PCB of tasks (struct rt_info) specific to real-time properties of tasks (worst-case execution time, deadline, …) [2]. Scheduling architecture in ChronOS is shown in Figure 2. In what follows, the implementation is explained.

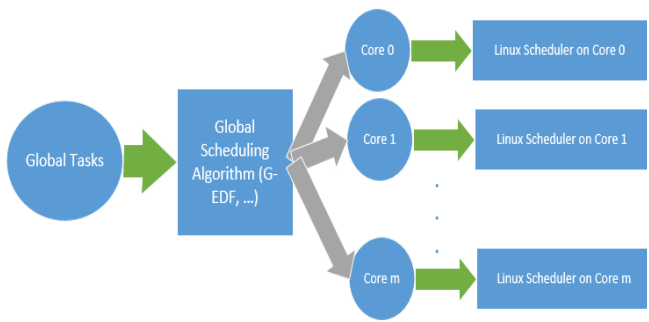

Figure 2. ChronOS global scheduling

### A. New Data Types

New data types added to ChronOS include new global priority queue (based on FC), which uses Linux kernel's mutex, as a global lock to underlying sequential data structure in FC and is a binary min-heap priority queue due to application in this work. This new types are shown below:

```
struct min_heap {
    int size;
    int capacity;
    struct rt_info **arr;
};
struct fc_queue {
    struct min_heap *priority_queue;
    struct mutex global_lock;
};
```

### B. Global Queue with FC Technique

According to FC technique as in [3], requests to global queue are handled as follows:

1- Each thread with a request, first writes it request to "publication list".

2- That thread tries to acquire global lock corresponding to FC.

3- If successful, traverse "publication list" and apply requests and return the results.

4- Else wait for the results (some other thread has the lock and is applying the requests).

After the above steps, every call in ChronOS, which results in a modification to global queue, including inserting a new task, removal of task and finding the task with the smallest deadline (NP-GEDF), is intercepted and is directed to corresponding FC methods:

PCB of the input task is modified to incorporate request type (insertion, removal,…) as explained in implementation of "publication list" in the next subsection. The above steps 1-4 are then executed.

### C. Modifications to ChronOS Data Types

One of the data types used in FC technique is "publication list" [3], which stores the concurrent requests to data structure. In order to facilitate this data type in ChronOS kernel, the following strategy is used: whenever a task thread has a new request, that request is written in the process control block (PCB) of the corresponding thread (task_struct in Linux kernel). Then during processing the request to data structure, thread which has the lock, traverses threads in the process and processes their requests. The following pseudo code depicts this strategy;

```
ProceessRequest(){
    for task t in process
```

```
        if t has a real-time request
                // no other thread in data
                // structure, so it's safe
                // modify it
                ExecuteReqOnPriorityQueue()
                Signal      completion    of
                request  to  thread  t  so  t
                can continue
}
```

In original FC, requests were written to thread local storage of the corresponding thread. Since this work is at kernel level, this strategy was not feasible anymore. Among the information stored in PCB, is the indication of being a real-time task and the request type (insertion, deletion, …).

### D. Integration with ChronOS

In order to use the new global queue with ChronOS some modifications were needed. Most notable ones are as follows:

1 – Originally in ChronOS, global queue was locked before any modification; this was not needed anymore.

2 – To check whether a given task is present in global queue or not, binary min-heap was traversed (instead of the original global linked-list).

### IV. EVALUATION

Planned experiments for implementation discussed is as follows: first we must devise a taskset such that it exhibits performance problems explained in section II (tasks have high amount of concurrency so there will be concurrent scheduling events). Then implementation will be tested on a CPU with a high number of cores. To demonstrate the performance gain, operations per second metric is going to be used. By using a concurrent data structure, processor cores will be better utilized and so there will be more operations performed in units of time, so by using this metric we can express improved concurrent performance in our program. Also another metric to be evaluated is worst latency, there could be a scenario during FC in which the execution time of thread performing combine would increase significantly as it has to execute requests on behalf of other threads as well it's own request.

### V. CONCLUSION

Efficient and scalable concurrent implementation of shared data structures is critical to performance in many applications on multicore platforms. One such application is real-time multicore scheduling and specifically G-NP-EDF scheduling algorithm. in this work one source of inefficiency in this algorithm was explained and an improved version based on FC technique was presented on ChronOS Linux kernel.

# Online Data-Driven Synchronization for IoT Systems with Unidirectional Networks

Seyed Hossein Hosseini Zahani
School of Electrical and Computer Engineering
University of Tehran
Tehran, Iran
hosseini.z@ut.ac.ir

Mehdi Kargahi
School of Electrical and Computer Engineering
University of Tehran
Tehran, Iran
kargahi@ut.ac.ir

*Abstract*—In this paper, we present an online method in unidirectional IoT systems to synchronize the data streams of multiple sensors, including wearable and environmental. Our proposed method is based on information theory concepts. First we use an entropy-based method to find events on environmental sensors data stream. After that, we use mutual information matching algorithm to find corresponding event in wearable sensors data stream. Indeed, we use physical interactions between wearable and environmental sensors in order to find shared events. Our experiments demonstrate that our proposed method can improve the quality of sensors data stream and therefore it can be used for the monitoring task in unidirectional IoT systems.

*Index Terms*—data-driven, synchronization, sensor, internet of things, unidirectional network, monitoring

## I. INTRODUCTION

With the advancement of technology, the number of sensors and electronic devices is increasing rapidly, and researchers predict that by 2020, there will be trillions of smart sensing devices in the world [1]. So the importance of IoT systems is growing. In some IoT applications such as smart home, understanding the relationship between multiple sensors can help us to obtain valuable information about the system. Understanding this relationship requires that the sensors have an accurate sense of timing between them. In other words, data of multiple sensors in such a system must be synchronized.

Typically, IoT systems have two-way communication between sensors and the central server, but in some cases which require high security, unidirectional networks are used. A unidirectional network is a network appliance or device allowing to transfer data only in one direction by imposing physical restrictions [2], [3]. In such networks only one-way communication with the central server is allowed . IoT devices will face a variety of network threats just like any other networked device. By using one-way communication setups it is possible to isolate IoT and other smart devices from remote attacks. An example of unidirectional network for prevention of exploit is shown in figure 1.

Many sensors such as low-power embedded devices can be mentioned in IoT systems that do not have an accurate internal real-time clock (RTC) or high-quality oscillators. So if sensor devices are not well-synchronized, collected data is less valuable.

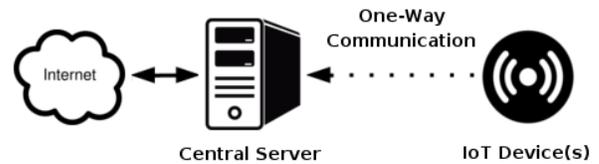

Fig. 1: IoT exploit prevention

Due to sensors in unidirectional networks unable to receive any packets from central servers, common time synchronization algorithms can not be used. Therefore we present an online method to synchronize data from multiple sensors for IoT systems with unidirectional network by using interaction of wearable and environmental sensors.

## II. RELATED WORKS

Many researches have been proposed in wireless sensor networks (WSN) due to the importance of time synchronization data in these networks. In [4] timing-sync protocol for sensor networks (TPSN) are developed. It works by establishing hierarchal networks and sending synchronization message to nodes in the network. Another important method for time synchronization in WSN is reference broadcast synchronization (RBS) method [5] in which nodes send reference beacons to other nodes that are near using physical-layer broadcasts.

Some methods are hardware based and they need to add components to the sensor devices in order to reduce clock drift. In [6], [7] they add some circuits such as recovery data and clock circuits to the sensors that generating special pulsed patterns and provide some extra information to synchronize the system.

There has been a lot of synchronization methods for WSN that based on two-way communication between sensors and central server. In [8] synchronization is done without centralized control. Sensors just exchange message with adjacent wireless nodes in two-way communication.

There are some offline methods such as [9]–[11] that synchronize data by finding coupling, alignment points and shared events. These methods are suitable for systems in which the sensors unable to have wireless communications due to power and other issues.

All the methods presented previously are not suitable for one-way networks. Our main goal in this article is to provide a solution for data synchronization in IoT systems with unidirectional network.

## III. BACKGROUND

Before discussing the synchronization method, we first discuss some of causes of clock inaccuracies. In an ideal IoT system, all sensor devices would have access to a high-accuracy clock such as global positioning system (GPS), global system for mobile communication (GSM) and etc. But if there is no possibility of receiving packets by sensor devices, synchronization method will be necessary.

Synchronization issues in sensor devices depend on the accuracy of hardware oscillators. To measure the accuracy of a hardware oscillator a measurement called oscillator stability has been proposed. The unit of oscillator stability measurement is typically parts per million (ppm), which is calculated as:

$$\text{ppm} = \frac{\Delta T}{T} \times 1,000,000, \quad (1)$$

Where $T$ is the total time passed and $\Delta T$ is delay or drift within $T$ (i.e. difference between the measured time passed and the actual time passed).

The oscillator stability value can vary according to the type of oscillator used in the sensor devices. For example for a crystal oscillator, the oscillator stability is within the range of $\pm 20 ppm$ to $\pm 5,000 ppm$. For other oscillators such as digitally controlled oscillators, voltage controlled oscillators, and relaxation oscillators this number is higher. The oscillators with higher accuracy usually are more expensive and consume more power [12]. Therefore, in many systems oscillators with low accuracy are used to reduce the cost and power consumption. so, the need for synchronization methods in these systems increases.

## IV. DATA-DRIVEN SYNCHRONIZATION

### A. System Model

Our system is an IoT system with unidirectional network. It consists of a number of environmental and wearable sensors and a central server. There is no communication between the sensors. Communication between the sensors and the central server is one-way. So, sensors can only send data to the server and the server can only receive data. Since checking online data can give us valuable information, monitoring incoming data on the central server is an essential task. In order to have reliable monitoring, we need to online synchronize the data.

We define set of all sensor devices in the system as:

$$S_n = \{s_1, s_2, \ldots, s_n\}, n \in \mathbb{N}, \quad (2)$$

where $n$ is the number of sensors. Each of the n sensors generates a data stream of tuples called observations. We denote each observation of data stream as $o_i^n$:

$$o_i^n = \{x_i^n, t_i^n\}, n \in \mathbb{N}. \quad (3)$$

Each observation contains a tuple of a data value and a corresponding timestamp that generated by local clock. $x_i^n$ is a data value at $i^{th}$ index of data stream of $n^{th}$ sensor and $t_i^n$ is corresponding timestamp, respectively.

**Definition IV.1** (shared event). A shared event is an event in the system that detected by multiple sensors.

From the global clock view a shared event has an exact time and multiple sensors can detect it without consideration of any local clock. We can use these shared events in order to synchronize data streams.

**Definition IV.2** (alignment point). An alignment point is a representation of a physical event in sensor data stream which is directly related to another sensor data stream and refers to the same event. We describe alignment point between two sensors as:

$$o_i^k \equiv o_i^l \quad \text{where} \quad k \neq l \quad (4)$$

When wearable sensors (e.g. wrist-worn sensors) having interaction with an environmental sensor (e.g. sensors on doors, cups, etc.), we can search in data stream of two sensors in order to find alignment point. If there is no clock drift in both sensor devices and they are synchronized at time 0, $t_i^k$ must be equal with $t_i^l$ but drift exist. Therefore, by finding all alignment points we can adjust their timestamp and synchronize data stream of sensors.

In order to detect alignment points we firstly search in environmental sensors to detect events because the environmental sensors are fixed unless they are affected by a person that worn wearable sensor. Secondly we have to search in data stream of wearable sensors to find the corresponding event. For detecting an event in environmental sensors we use an entropy-based method and for finding corresponding event in another sensors (i.e. wearable sensors) we use a matching algorithm based on mutual information theory.

### B. Entropy-Based Event Detection

Entropy is defined as the average amount of information in a signal [13]. In other words, it is a measurement of the randomness of a signal and calculated as

$$H(X) = -\sum_{i=1}^{n} p(x_i) \log_2 p(x_i), \quad (5)$$

where $X$ is a discrete random variable with possible values $\{x_1, \ldots, x_n\}$ and probability mass function $p(x)$. In our work, $p(x)$ is the probability of a given $x$ in the signal's distribution and it is calculated based on histogram.

In Figure 2 an example of entropy calculations for a raw data of an arbitrary accelerometer is shown. As you can see, the amount of entropy in the segment that the accelerometer has some motions is significant, which is what we are looking for.

So, by choosing an sliding window, moving it through the data stream of environmental sensors and finding segments with high entropy values, we can detect events. When an event detected in first sensor data stream (i.e. environmental sensors)

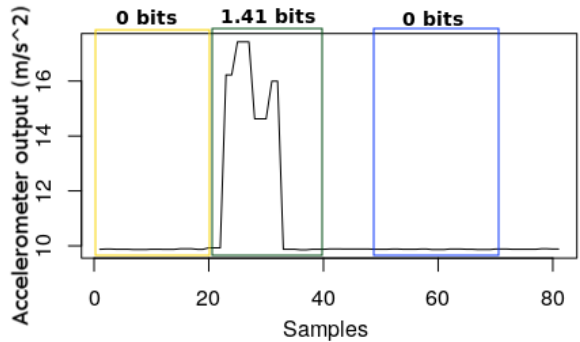

Fig. 2: Entropy calculations for various segments in an accelerometer data stream

we should match it in second data stream (i.e. wearable sensors) in order to find shared event.

*C. Mutual Information Matching Algorithm*

Mutual information is a matching algorithm to measure the amount of information shared between two signals [14] and is calculated as:

$$I(X;Y) = \sum_{y \in Y} \sum_{x \in X} p(x,y) \log_2 \left( \frac{p(x,y)}{p(x)p(y)} \right), \quad (6)$$

where $X$ and $Y$ are two discrete random variables, $p(x,y)$ is the joint probability function of $X$ and $Y$, and $p(x)$ and $p(y)$ are the marginal probability distribution functions of $X$ and $Y$ respectively. In our work, $X$ and $Y$ represent two signals.

After an event is founded in first sensor data stream (i.e. environmental sensor) through entropy method, we should determine which wearable sensor data stream is associated with it. For this purpose, we define a search window for second data stream based on index of first data stream:

$$search\_window = t_i^k \pm (t_i^k \times \frac{ppm_l}{1,000,000}), \quad (7)$$

where $t_i^k$ is the timestamp of alignment point in first data stream, and $ppm_l$ is the stability of second sensor.

After that, we calculate mutual information between first data stream and all wearable sensors data stream. The mutual information is performed only in the $search\_window$ of each wearable sensors. Through moving an sliding window on second data stream we can find peak of mutual information calculations. Then we compare all peaks and select the wearable sensor that has the maximum value. With this method, we can determine which of the wearable sensors are related to first data stream (i.e. environmental sensor). Finally we adjust the timestamp of alignment points in order to synchronize their data.

## V. EXPERIMENTS AND RESULTS

*A. Experimental Setup*

We have not yet managed to complete the full experiments. In our initial experiment, we conducted experiment offline and just worked on two sensors, one of which is a wrist-worn sensor and the other is an environmental sensor that attached to a coffee mug. For this purpose, we used two smartphones with 3-axis accelerometer sensor. We collected data at 20Hz and stored on local SD card. This experiment lasted 30 minutes. R [15] was used to process experimental data.

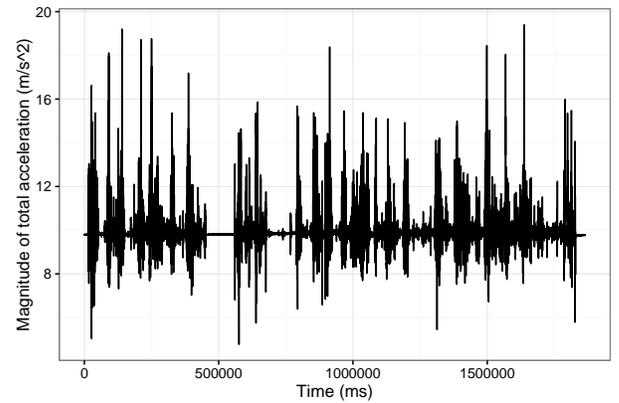

(a) Wrist-worn sensor

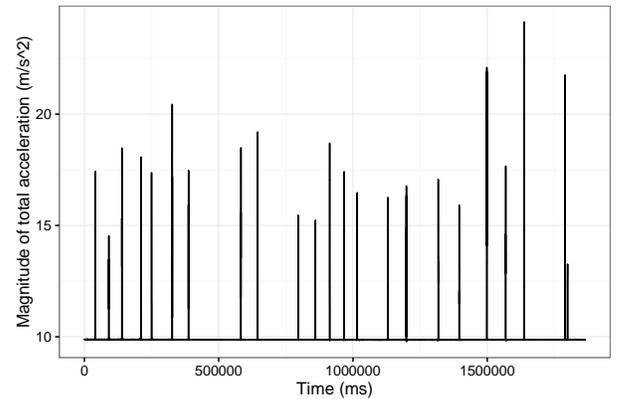

(b) Environmental sensor

Fig. 3: Original collected data

Original data is shown in figure 3 for both wearable (3a) and environmental (3b) sensors. As you can see in figure 3b, there are 23 shared events that should be detected.

Smartphones usually are synchronized with NTP protocol so timestamps of collected data was correct and there was no drift (figure 3). In order to had a low accuracy oscillator conditions, we needed to simulate the oscillator stability (ppm). So we simulated 6,000 ppm for wrist-worn sensor and 5,000 ppm for environmental sensor using random number generator techniques.

*B. Results and Analysis*

By using entropy-based method on environmental sensor, we could detect all 23 shared events well. After that, we had to find the corresponding event in the second sensor (wrist-worn). For each detected event, we selected a search window on second sensor and by moving an sliding window through second data stream we calculated mutual information values.

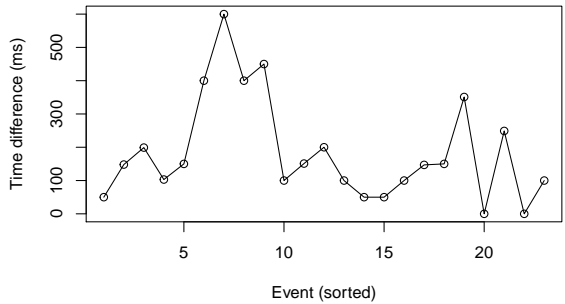

Fig. 4: Time differences of detected events in the first sensor and corresponding events in the second sensor

As we already said we selected peak of mutual information outputs.

In figure 4 you can see time differences between detected events in the environmental sensor and corresponding events in wrist-worn sensor. As shown in plot, maximum difference time is 600 millisecond and it is related to 7th event. Ideally and without drifts, time differences of all events must be zero.

In the future, we need to focus more on the accuracy of our proposed method. We should define a more appropriate $search\_window$. We should use more sensors in our experiment. Also we have to increase the duration of experiment. Finally we have to test our method in online mode while data is coming from sensor devices.

# Semi-partitioned Scheduling Hard Real-time Periodic DAGs in Multicores

Mojtaba Hatami
Department of Computer Engineering
Ferdowsi University of Mashhad, Mashhad, Iran
hatami.mojtaba@mail.um.ac.ir

Mahmoud Naghibzadeh
Department of Computer Engineering
Ferdowsi University of Mashhad, Mashhad, Iran
naghibzadeh@um.ac.ir

*Abstract*—Recent trends in real-time systems are towards multicores and parallel processes in the form of directed acyclic graphs. The scheduling aspect of such systems has been worked on and many methods are developed. Nevertheless, the need for more efficient approaches which can use fewer number of cores has not vanished. Semi-partitioned scheduling of hard real-time parallel tasks in multicores is studied in this paper. Since there is no benefit in completing a task much before its deadline, after scheduling a parallel task, if it is beneficial for other tasks, the execution of this task is further moved towards its deadline, i.e., stretching, to make room for tasks with closer deadlines. A new concept, prior+, load of tasks is used to rank all tasks of each directed acyclic graph and order them for scheduling. The scheduler is offline and the schedule map is used during run time. One benefit of this is the reduction of the scheduling overhead during run time which helps to safely accept loads. The comparative evaluations show the algorithms performance is superior to the state of the art ones. It also confirmed that the new concept of prior+ load of a task is very effective in scheduling real-time directed acyclic graphs and suggests that it can as well be useful in scheduling workflows.

*Index Terms*—Real-time parallel tasks, hard deadlines, semi-partitioning, multicore processors, offline scheduling

## I. INTRODUCTION

Advances in hardware technology has made production of multicore processors on one chip possible. A great advantage of such processors is their capability of running parallel tasks much faster than running their sequential versions. Besides, real-time processes of some systems have become more complex and they are each transformed from a sequential task into many interconnected tasks with precedence constraints. A common model of these dependent tasks is Directed Acyclic Graphs (DAG). Although there are precedence relations between these tasks, many of them can run in parallel. In this paper, DAG and parallel task are used interchangeably.

To make use of the new technologies, multicores and DAGs of tasks, we need to develop new scheduling strategies to better utilize the hardware resources and improve the quality of services needed by the applications. Although many schedulers are developed for sequential tasks to run in the multicore systems, their results cannot simply be extended to DAGs of tasks. As an example of a multifaceted real-time system composed of many complex tasks consider the control subsystem of autonomous cars. Navigation path finder, object recognition, speed limit recognition, machine vision, and intelligent decision making are samples of complex real-time tasks [1]. Safely running all such complex tasks on a single processor is very difficult, if not impossible.

A restriction present in some of the previous work is the assumption that tasks are preemptible. This assumption makes the safety analysis of the system simpler because often the least upper bound to utilization that is developed by Liu and Layland [2] can be used for each core, for the case of segmentation. However, runtime scheduling overhead and context switching are two drawbacks of this assumption. Each context switch may make the cached data of the outgoing task invalid for the next time it is resumed. Reload of the removed data takes time and increases the tasks actual execution time. It can even make the execution time estimation of the tasks uncertain. With DAG of tasks, the granularity of tasks becomes smaller and the need for preemption to achieve higher utilization becomes unnecessary.

In this paper, scheduling multiple periodic DAGs of tasks in multicores is investigated. The scheduler makes use of each tasks prior+ load as its priority to be selected for execution. Prior+ load of a task is defined to be the total execution time of all its fathers and its ancestors plus its own execution time. Important contributions of this paper are: Introduction of the new ranking method based on prior+, Development of a new list-based scheduling algorithm using the stretching concept, backward filling of time periods of cores, and bottom-up scheduling of tasks of every DAG by starting from the leaves of a DAG and going toward its beginning.

## II. RELATED WORK

Scheduling algorithms of sequential periodic tasks in multicore systems are in two categories, global and partitioned [3]. In the global model, each request of a task, as a whole, may be scheduled on any available core. In the partitioned algorithms, each task is assigned to a specific core and all its requests are executed by the same core. There is a hybrid approach called semi-partitioned in which some tasks are each completely assigned to a specific core and some tasks are broken into subtasks and each subtask is assigned to a different core [4]. Semi-partitioned methods introduce new overheads and new complexity to scheduling, while supporting higher utilization degree, hence, this kind of methods often use fewer number of cores [4]. Another model of real-time periodic tasks is the parallel model. Several task models have been proposed for modeling parallel tasks [1], [5]. Most famous of which are the Gang Model [6], DAG Model [1], [7], and fork-join Model [8]. Reference [1] studies a parallel task model in which a segmentation approach is used and each segment is composed of an arbitrary number of threads. After segmentation, tasks are considered to be independent, hence well-known algorithms such as RM and EDF can be used to schedule these tasks.

## III. TASK MODEL

The Set $\tau = \{\tau_1, \tau_2, \ldots, \tau_n\}$ contains $n$ DAGs, in which each node of a DAG represents a non-preemptable task and the edges between the nodes represent the dependency relationship between them. The paper by Buttazzo [9] states that each of preemptive and non-preemptive approach has advantages and disadvantages, and no one dominates the other. The specific reason for us to use the non-preemptive version is that by breaking a sequential real-time task into a DAG of tasks to run them in parallel (when possible), the new tasks are smaller with respect to execution times needed. Besides, with the advances in the speed of processors many more instructions can be executed in one time unit. From now on, $\tau_i$ is defined to be the i-th DAG of tasks in the DAG set. We also use task and node terms interchangeably.

Each node in the DAG $\tau_i, i: 1, 2, \ldots, n$, is defined as $w_i^j$ : $1 \leq j \leq n_i$ where $n_i$ denotes the number of DAGs nodes in the DAG $\tau_i$. The maximum execution time of each task in this DAG is represented by $E_i^j$. A $w_i^j \rightarrow w_i^k$ directed edge indicates that $w_i^k$ should wait until the end of the execution of $w_i^j$. In this case, the $w_i^j$ and $w_i^k$ are called the parent and the child, respectively. A task can only start running when the execution of all its parents has completed. The period of the repetition of each DAG $\tau_i$ is represented by $T_i$. The deadline parameter ($D_i$) of each $\tau_i$ is implicit and it is equal to $T_i$. The execution time to run a DAG ($C_i$) is equal to the sum of execution times of its tasks. Therefore, $C_i$ is equal to the maximum execution time of $\tau_i$ on a single processor with unit speed. For a DAG $\tau_i$, the critical path ($CP_i$) is a directed path that has the highest execution time among all the directed paths in the DAG $\tau_i$.

## IV. MOTIVATION EXAMPLE

Consider the DAGs of tasks in Fig.1(a),(b). Suppose that the request periods (also deadlines) for DAGs 1 and 2 are 20 and 10, respectively. If the number of available cores to running such periodic parallel tasks is three, the Global EDF (GEDF) algorithm and the Decompose (which we will call DEC from here on) algorithm presented in [1] are not able to schedule these DAGs. In fact, for safe scheduling, the GEDF algorithm needs four cores and the DEC algorithm needs five cores. Using the method presented in this paper a valid schedule for these two DAGs of tasks shown in Fig.1(c). Since the Least Common Multiple (LCM) of DAGs is 20, the schedule map is produceed for this period for which the first DAG is scheduled twice while the second one is scheduled once. It is clear that an

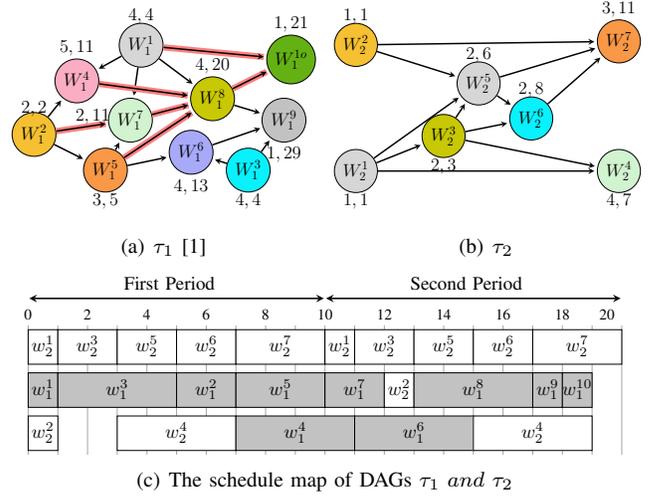

(a) $\tau_1$ [1]   (b) $\tau_2$

(c) The schedule map of DAGs $\tau_1$ and $\tau_2$

**Fig. 1:** DAG set $\tau = \{\tau_1, \tau_2\}$ and The schedule map

example does not prove its superiority but a hope to improve the state of the art.

## V. THE PROPOSED ALGORITHM

The proposed algorithm is an offline one and, as opposed to the traditional approach, it starts scheduling each DAG from the exit task(s) and continues toward the entry task(s). Obviously, the exit task(s) are scheduled close to the deadline of the task graph and their parents are scheduled further from the deadline and so on. The strategy is a semi-partitioned one in the sense that most cores are each dedicated to tasks of one DAG and few other cores each have tasks of more than one DAGs. Semi-partitioned have been commonly used for simple, i.e., non-DAG, tasks [3]. One other property of the proposed algorithm is that the schedule map is produced for the LCM of periods of DAGs. If such a schedule is feasible, since the same schedule is repeated from then on, the system will be safe and there will be no task misses. The whole scheduling process is composed of five steps, (1) task ranking of each DAG, (2) estimating the number of required cores for each DAG, (3) producing a primary schedule for each DAG, (4) compacting the schedule maps and (5) extending the primary schedule maps for the LCM period.

### A. Task ranking based on Prior+

For each DAG, all tasks are ranked first. The ranking approach here is novel and different from what is usually called upward rank [10]. The central concept behind this new ranking system is prior+ load of each task, which is defined to be the total execution time of all its fathers and its ancestors plus its own. To illustrate the concept of the prior+ load of nodes, Fig.1(a) is used. The numbers on the nodes, from left to right, represent the nodes execution time and its prior+ load. For example, the prior+ load of node $w_1^{10}$ is equal to the execution time of all nodes connected by colored edges, which is 21. To prioritize the nodes of a DAG, a node with a higher prior+ is given a higher priority than that of a lower

prior+ value. If two nodes have the same prior+ value, the node with smaller execution time gets a higher priority. If the prior+ load and the execution time of two nodes are the same, one of them is randomly assigned a higher priority. Therefore, the descending order of priorities of the nodes in the DAG of Fig.1(a), from left to right is: $w_1^9$, $w_1^{10}$, $w_1^8$, $w_1^6$, $w_1^7$, $w_1^4$, $w_1^5$, $w_1^3$, $w_1^1$, $w_1^2$.

### B. The number of resources needed to schedule a DAG

Suppose that the frequency, or speed, of the processor for calculation of tasks execution times is of Type1. For the prediction of the number of processors to safely run a DAG of tasks, we first calculate the Earliest Start Time (EST) and the Latest Finish Time (LFT) of all tasks of the DAG. These concepts have been used in workflow scheduling (represented as DAGs or Hybrid DAGs) [10]).

In the next step, tasks are clustered. All tasks on the critical path are put in a separate cluster and all other tasks are clustered such that all tasks with the same EST form a separate cluster. Afterwards, the density of each cluster, i.e., the sum of execution times of all tasks of the cluster divided by the maximum difference of the LFT and the EST of all tasks of the cluster, is calculated. The ceiling of density value is the estimated minimum number of required processors to schedule the DAG.

### C. A primary schedule for each DAG

To create a primary schedule map of a DAG, the algorithm uses the minimum number of resources that was calculated in the previous subsection. If the minimum number of resources is more than available cores, we temporarily assume we have enough cores because a compaction process will follow. The final decision on the schedulability of all DAGs will be made after compaction. If the estimated number of cores is not enough, cores are used one at a time until the DAG is scheduled. Algorithm 1 is developed to schedule a DAG and create a schedule map. The map, $mp$, is for only one period of the DAG. In this algorithm, variable $Q$ denotes a priority queue (based on prior+) of tasks. DLFT (Dynamic LFT) is used in Line 9 and shows the latest time by which the execution of a task, considering the execution of its children, is completed, Formula (1). Recall that the scheduling starts from the leaves of a DAG and continues toward its beginning. If children of a task are later moved from one core to the other, this variable will change.

$$DLFT(w_i^j) = \begin{cases} LFT(w_i^j), & w_i^j \text{ is exit nod} \\ \min(DLFT(w_i^k) - E_i^k), & \text{else} \\ w_i^k \in child(w_i^j) \end{cases} \quad (1)$$

Line 3 of Algorithm 1 calculates the number of processors needed to schedule each DAG. The FM in Line 5 shows the Final Moment of free interval of a core. It means that no task has been scheduled on either cores for the time interval [0, FM]. In Line 6, all DAG nodes that do not have any child (exit nodes) are added to the $Q$. After the scheduling of each task, all parents of that task whose all children are scheduled, are added to the $Q$. With DAG of Fig. 1(a), Algorithm 1 proceeds from left to right, as: $w_1^9$, $w_1^{10}$, $w_1^8$, $w_1^6$, $w_1^7$, $w_1^4$, $w_1^5$, $w_1^3$, $w_1^1$, $w_1^2$. A selected task for scheduling, is scheduled on a core that maximizes the $Min(DLFT(w), FM(core_r))$. The finish time and start time of the task are calculated and added to mp of $Core_r$.

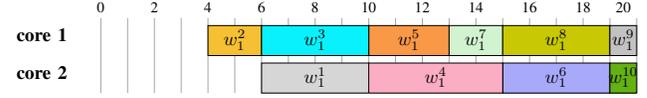

**Fig. 2:** Schedule map of core 1 and 2

---

**Algorithm 1** *Primary Scheling*

**Input:** $\tau$
**Output:** $MP = \{mp_1, mp_2, mp_3, , mp_x\}$
  /* $mp_k$ is scheduling map for k-th core/*
1: **while** All $\tau_i \in \tau$ not scheduled **do**
2:   calculate $prior + load$ of all $w_i^j \in \tau_i$
3:   $M \leftarrow$ the minimum number of cores needed for scheduling requirement $\tau_i$.
4:   $Reserve$ these M cores for scheduling $\tau_i$.
5:   $FM \leftarrow D_i$ for each reserved core
6:   Add all exit node to Q
7:   **while** $Q$ is not empty **do**
8:     $W \leftarrow$ Dequee(Q)
9:     schedule W on the $mp_r$ that $core_r$ maximize $\alpha$ =min(DLFT(W), FM($Core_r$))
10:    Finish Time(W) $\leftarrow \alpha$
11:    Start Time(W) $\leftarrow \alpha$ - Execution Time(W)
12:    FM($core_r$)$\leftarrow$ Start Time(W)
13:    Add all parent of W to Q if all of its children is scheduled
14:   **end while**
15:   Compact the new scheduling maps
16: **end while**
17: **return** $theMP$

---

### D. Compacting the schedule maps

The compaction of the schedule map uses two integers $Aindex$ and $Bindex$ representing the first and the last core to participate in the compaction. The compaction is done on schedule maps of cores Aindex, $Aindex + 1, \ldots, Bindex$. A schedule map $mp_i$ is selected and for each task of the map the algorithm tries to fill the empty gap, if any, just prior to this task with other tasks in the $mp_i$ set. For this purpose, tasks with lower ranks have higher priorities. The algorithm follows a best fit method to fill the unassigned gaps with minimum penalty. Penalty is defined to be the length of time from the start of the gap to the start of the newly assigned task to the gap. With respect to this objective Task $w_i^j$ is preferred over Task $w_k^h$ if either task $w_i^j$ has a lower rank than $w_k^h$ or relations (3) and (4) hold, simultaneously. For these equations, DEST is defined as (2):

$$DEST(w_i^j) = \begin{cases} EST(w_i^j), & w_i^j \text{ is entry nod} \\ \max(DEST(w_i^k) + W_i^k), & \text{else} \\ w_i^k \in parents(w_i^j) \end{cases} \quad (2)$$

$$DEST(W_i^j) + E_i^j \leq DEST(W_i^h) + E_i^h \quad (3)$$

$$DEST(W_i^j) - \max(DEST(W_i^j), GapStart) \leq$$
$$DEST(W_i^h) - \max(DEST(W_i^h), GapStart) \quad (4)$$

The inequality (3) checks whether the finish time of task is earlier than the finish time of task $w_k^h$. The inequality (4) checks whether the penalty for Task $w_i^j$ is less than penalty for Task wkh .To compact the schedule maps of Fig.2, first, the task $w_1^2$ is selected as temp and the range [0, 4] is considered as the gap. Then, it is examined whether it can find a node other than $w_1^2$ that can run in this gap (node with least priority). In this case, the response is negative, so $w_1^2$ is selected and re-scheduled (Fig.3(a)). The next Task, $w_1^1$, is chosen as temp, while the interval [2, 6] is the gap and the search is performed again. Now Core 1 is fully utilized. Finally, the compacted schedule maps will be as depicted in Fig.3(b).

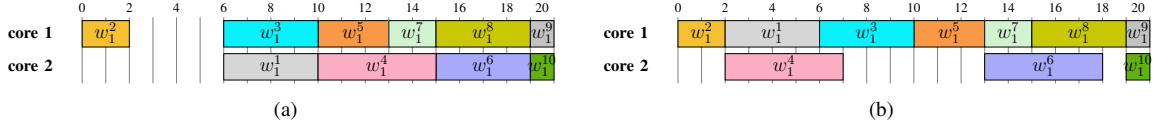

Fig. 3: Compacting the schedule map of Cores 1 and 2

*E. Extending the primary schedule map*

Since DAGs are periodic, in this step, for each DAG $\tau_i$, we repeat the primary schedule map $t$ times where $t$ is calculated as(5):

$$t = \frac{LCM(D_1, D_2, \ldots, D_n)}{D_i} - 1 \qquad (5)$$

Afterwards, we add $D_i * k$ to the start and finish time of all nodes in the k-th copy. The prior+ of each task $w_i^j$ in the k-th copy will be updated and will become equal to $Prior^+(w_i^j) + C_i * k$.

*F. The proposed main algorithm*

The proposed main algorithm takes a set of DAGs and $m$ as the number of CPU cores. It performs Algorithm 1 as part of the method. In summary, if the number of cores that are required to schedule all tasks is at the most $m$, the scheduling is successful, otherwise, the scheduling is unsuccessful.

**Algorithm 2** the proposed algorithm

**Input:** $\tau = \{\tau_1, \tau_2, \ldots, \tau_n\}$, m
**Output:** $MP$
1: $MP \leftarrow$ Create Primary Scheduling map : Algorithm1 $(\tau)$
2: Extend the primary scheduling map's
3: Compacted the MP
   /*Aindex=1, Bindex= x, x is the number of the reserved cores/*
4: **if** there are enough cores for all DAGs **then**
5:     **return** $Compacted\ MP$
6: **else**
7:     **return** $Scheduling\ Failed.$
8: **end if**

## VI. EVALUATION

We did a comprehensive set of comparative simulations but due to space limitation only reference [1] is selected. Their algorithm decomposes the graphs and then uses the Global EDF algorithm to schedule the DAGs. Since the method presented here is non-preemptive, the comparison is done using the non-preemptive version of DEC algorithm [1]. DEC was described without any ambiguity and we implemented its non-preemptive version in the same environment as our proposed algorithm.

In all experiments, every collection of tasks five DAGs. In the first experiment, 1000 collections of DAGs were generated with p=0.6 (p is the probability of adding an edge between two nodes). Both algorithms were used to schedule each collection using 4, 8, 12, 16, and 32 cores systems. The results are shown in Fig.4(a). The success rates of the proposed algorithm is higher than that of DEC. We also compared the algorithms for utilization of used cores and ours was superior, Fig.4(b).

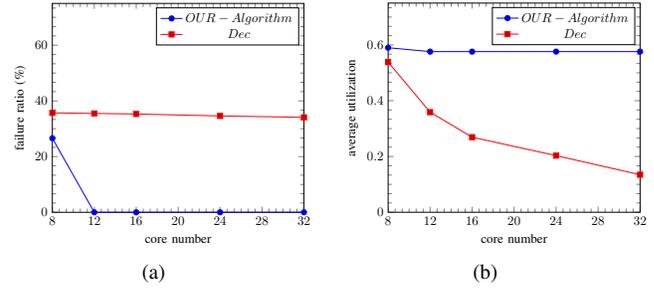

Fig. 4: Experiment result with different number of cores

**Conclusions:** In this paper, we consider scheduling sets of DAGs and we proposed the prior+ load concept and a multi-level semi-partitioned scheduling algorithm. It is an offline one and, as opposed to the traditional approaches, it starts scheduling each DAG from the exit task(s) and continues toward the entry task(s). We examined the performance of the proposed algorithm through simulation, and we have observed that the success ratio of the algorithm is superior to competitive approaches.

# Energy Debugging of Android Applications based on User Behavior

Boshra Taheri  
School of Electrical and Computer Engineering  
University of Tehran  
Tehran, Iran  
b.taheri@ut.ac.ir

Fathiyeh Faghih  
School of Electrical and Computer Engineering  
University of Tehran  
Tehran, Iran  
f.faghih@ut.ac.ir

Mehdi Kargahi  
School of Electrical and Computer Engineering  
University of Tehran  
Tehran, Iran  
kargahi@ut.ac.ir

*Abstract*—In recent years, Android have become a popular operating system for smartphones. Regarding this growing trend, a wide variety of applications are being developed for this platform. Due to limited energy supply on such smartphones, the energy consumption of an application plays a significant role in users' satisfaction. Recent studies in this field reveals that a considerable number of applications suffer from energy issues. To this end, we are going to provide a systematic framework for application developers aiming to improve the level of energy consumption in application. In this research, we elaborate a profiling technique in order to model users' behavior in accordance with how they use the application.

*Keywords—Android; Energy Bug; Energy Hotspot; Energy Consumption; User Behavior*

I. INTRODUCTION

Considering the use of Android-based smartphones and the increased variety and complexity of applications provided to them, energy consumption issues have become particularly important. Given the fact that these devices are battery-based, and therefore the amount of energy available to them is limited, the amount of energy consumed by applications is a major factor in satisfying their users. On the other hand, the growth of battery-related technologies is not synchronous with the pace of increase in the complexity of applications. Therefore, it is necessary to analyze and optimize the energy consumption in Android applications.

According to [1], *energy bug* is an error in the system that causes the system to consume more energy than expected. Its source can be the application, operating system, hardware, firmware, or environmental factors. The application developer cannot solve the energy problems associated with the operating system, hardware, or firmware. However, he can address the energy problems and eliminate them or reduce their impact by applying changes at the level of architecture or design of the application program or its source code.

In a smartphone device using an OS like Android or iOS, there are a variety of hardware components, such as a GPS receiver or wireless network card, that use energy due to their interactions with applications. Some of these components cause more energy consumption than others. In order to use a hardware component that is idle or off, it is necessary for the application to use system calls causing the hardware component to go into its active state, in which its power consumption will be higher. When the hardware component is not needed, the application is expected to execute another system call and return the hardware component to its base state. One example of energy bugs is not releasing the resources after their usage is complete, which is called "resource leak" in the literature. Energy bugs, such as resource leak do not affect the functionality of the application but result in a lot of energy consumption after the program terminates, resulting in shorter battery life for the smartphone.

Another type of energy bug, called "energy hotspot" causes excessive energy consumption during the program execution." An example of this kind of problems is the low-level utilization of a hardware component, which happens when an application activates a component earlier than the time it is needed or turns it off or idle later than it can. Another instance of this type of bug can occur when an application requests to receive a sensor's updates. In this case, the application must determine the intervals at which the updates are provided. The smaller the period, the more accurate the information it receives from the sensor.

In case of sensors, one of the factors affecting the optimum value for the update periods is user behavior. For example, in the case of a GPS receiver, if the user is on the move, it is better to report its location information at a higher rate to the application, but if he stays at a location, an increase in the interval between two location updates does not cause a disruption to the functionality of the application. Therefore, studying users' behavior can lead to decisions resulting in lower energy consumption, while the application functionality is also preserved. For example, consider an e-hailing service, and assume that studying users' behavior shows that many users are stationary at the time of taxi order and during the waiting time for a taxi, and their location does not change. In this scenario, the developer can reduce the location update rate during the waiting time and improve the energy consumption of the application. In this research, our goal is to propose a method that can help the designer or developer to improve the application from the energy perspective, considering the users' behavior. The input of this method is the source code of the application program, along with the activity logs of the users of the program, and the outputs are suggestions to the designer or developer to improve the energy consumption of the application program by more efficient usage of the hardware components, such as the GPS receiver.

## II. PROBLEM STATEMENT

Recent studies on energy usage of Android applications show that a significant number of applications still suffer from energy inefficiencies, which has a significant impact on users' satisfaction [2], [3]. Researchers have recently begun to study this problem from the application perspective and have proposed innovative methods to debug Android applications considering the energy usage [2]–[4]. However, our studies show that there is a gap in the research in this area, that is considering the users' behavior when debugging the applications. As mentioned before, users' behavior can have a significant impact on how we can improve the energy consumption of an Android application.

So, the idea in this research is to propose a systematic method that can help application developers improve the energy consumption of Android applications, considering the users' behavior. Since each suggestion for improvement is based on the behavior of a sample of users of the program and cannot be generalized to all users, it is possible for the developer to create different profiles based on the given suggestions, and select one of them to run at run-time, based on certain parameters.

## III. PROPOSED FRAMEWORK

Figure 1 shows the overview of the proposed method. The method will be such that at first, the repository is empty, and the application program is the same for all users and in all conditions. In the first step, the application program is instrumented with a tool such as soot [5] to record logs of certain methods of Android framework. After installing the program on the user's handset, the log of the user's interaction with the application is recorded and stored for a certain period of time. Then these logs, along with the application program model and the power model of the hardware components, are given as inputs to the analyzer module. The application model can be created using a tool such as FlowDroid [6]. The power model of hardware components can be obtained from their specifications sheet. The main component in this research is GPS receiver, as it is one of the most energy-hungry hardware components in a smartphone device. In the analyzer module, the proposed algorithms are implemented on the input and in the output, it returns suggestions for improvement that are added to the profile repository. The analyzer module will use methods or tools for estimating energy consumption at the code level. It may also require parameters from an application program to perform the analysis, which must be specified by the designer or developer. For example, if an application uses a GPS receiver, it may be necessary to obtain the program's required precision for the location from its designer or developer.

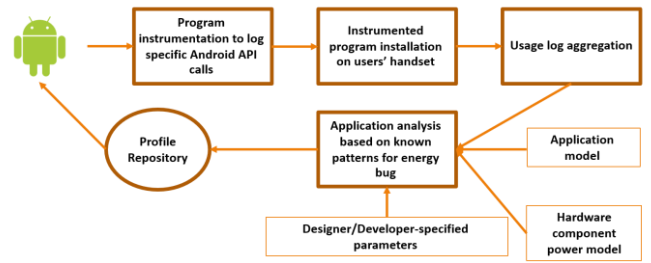

**Figure 1 The overview of the proposed method**

Finally, this approach will be tested on a number of open source applications that provide location-based services and use GPS receiver as the location provider, and the impact of the suggestions on their energy consumption will be evaluated.

## IV. RELATED WORK

In the literature, there are tools or methods for identifying and solving systemic energy problems. In each of these studies, a particular type of problem has been addressed. The work presented in [3] provides a framework that can detect sources of resource leak problems for sensors as well as Wake Lock. This framework can also detect energy hotspots regarding sensor data. Energy hotspots for the sensor, based on what is presented in this article, are equivalent to the low productivity of the data reported by the sensor. In this research, a metric for measuring the efficiency of sensor data is proposed and bounded symbolic execution is used as the dynamic code analysis method. [4] addresses forms of resource leak and provides a framework that can detect such energy problems using a static analysis method and unlimited symbolic execution, which is a dynamic code analysis method. This framework is also able to offer patches to fix the identified problems. In these works, user behavior is not considered while we believe that software energy consumption pattern can be improved with regard to the pattern of its usage.

## V. CONCLUSION

Energy consumption of an application is an important factor in user satisfaction. So, it is important to develop tools and methods to help developers evaluate their applications from this perspective. In this research, we propose a systematic method to help android application developers to improve their applications' energy consumption with regard to their users' behavior.